\documentclass[prl,twocolumn,aps,showpacs]{revtex4}

\usepackage{here}
\usepackage{graphicx}
\usepackage{bm}

\newcommand{\be}{\begin{equation}}
\newcommand{\ee}{\end{equation}}

\newcommand\pictc[5]{\begin{figure}[t]
                       \centerline{\vspace{-0mm}
\includegraphics[width=#1\columnwidth,height=0.7\textheight,keepaspectratio]{#3}}
                       \protect\caption{\protect\label{fig:#4} #5}\vspace{-2mm}
                    \end{figure}            }
\newcommand\pict[4][1]{\pictc{#1}{!tb}{#2}{#3}{#4}}
\newcommand\rpict[1]{\ref{fig:#1}}

\newcounter{Fig}

\begin{document}
\begin{sloppy}

\title{Observation of surface gap solitons in semi-infinite waveguide arrays}

\author{Christian R. Rosberg$^1$, Dragomir N. Neshev$^1$, Wieslaw Krolikowski$^2$, Arnan~Mitchell$^3$, Rodrigo A. Vicencio$^4$, Mario I. Molina$^5$, and Yuri S. Kivshar$^1$}
\affiliation{$^1$Nonlinear Physics Centre and $^2$Laser Physics Centre, Centre for Ultrahigh-bandwidth Devices for Optical Systems (CUDOS),
Research School of Physical Sciences and Engineering, Australian National University, Canberra, Australia\\
$^3$School of Electrical and Computer Systems Engineering, RMIT University, Melbourne, Australia\\
$^4$Max-Planck-Institut fur Physik komplexer Systeme, Dresden, Germany \\
$^5$Departamento de F\'{\i}sica, Facultad de Ciencias, Universidad
de Chile, Santiago, Chile}
\pacs{42.65.Jx, 42.65.Tg, 42.65.Sf}

\begin{abstract}
We report on the first observation of {\em surface gap solitons}, recently predicted to exist at the interface between uniform and periodic dielectric media with defocusing nonlinearity [Ya.~V. Kartashov {\em et al.}, Phys. Rev. Lett. {\bf 96}, 073901 (2006)]. We demonstrate strong self-trapping at the edge of a LiNbO$_3$ waveguide array and the formation of staggered surface solitons with propagation constant inside the first photonic band gap. We study the crossover between linear repulsion and nonlinear attraction at the surface, revealing the mechanism of nonlinearity-mediated stabilization of the surface gap modes.
\end{abstract}

\pacs{42.65.Tg, 42.65.Sf5, 42.65.Wi}

\maketitle

\noindent

Interfaces between different physical media can support a special class of localized waves known as surface waves or surface modes. In periodic systems, {\em staggered surface modes} are often referred to as Tamm states~\cite{Tamm:1932-849:ZPhys}, first identified as localized electronic states at the edge of a truncated periodic potential. Because of the difficulties in observing this type of surface waves in natural materials such as crystals, successful efforts were made to demonstrate their existence in nano-engineered periodic structures or superlattices~\cite{Ohno:1990:PRL}. An optical analog of linear Tamm states has been described theoretically and demonstrated experimentally for an interface separating periodic and homogeneous dielectric media~\cite{yariv,Yeh_APL_78}.

Nonlinear surface waves have been studied in different fields of physics and most extensively in optics where surface TE and TM modes were predicted and analyzed for the interfaces between two different homogeneous nonlinear dielectric media~\cite{tomlinson,book,mihalache}. In addition, nonlinear effects have been shown to stabilize surface waves in discrete systems, generating different types of modes localized at and near the surface~\cite{physica_d}. Self-trapping of light near the boundary of a {\em self-focusing} photonic lattice has recently been predicted theoretically~\cite{Makris:2005-2466:OL} and demonstrated in experiment~\cite{Suntsov:2006-063901:PRL} through the formation of {\em discrete surface solitons} at the edge of a waveguide array.

Recently, Kartashov {\em et al.}~\cite{Kartashov:2006-073901:PRL} predicted theoretically the existence of {\em surface gap solitons} at the interface between a uniform medium and a photonic lattice with {\em defocusing} nonlinearity. In such systems, light localization occurs inside a photonic bandgap in the form of staggered surface modes. This enables us to draw an analogy with the localized electronic Tamm states and extend it to the nonlinear regime, so that the surface gap solitons can be termed as {\em nonlinear Tamm states}. They posses a unique combination of properties related to both electronic and optical surface waves and discrete optical gap solitons. The ability to generate such surface gap solitons could provide novel and effective experimental tools for the study of nonlinear effects near surfaces with possible applications in optical sensing and switching.

In this Letter we study experimentally self-action of a narrow beam propagating near the edge of a LiNbO$_3$ waveguide array with defocusing nonlinearity. For the first time to our knowledge, we observe the formation of surface gap solitons, or nonlinear Tamm states. While linear surface modes {\em do not exist} in this type of system, discrete light self-trapping is observed in the nonlinear regime above a certain threshold power when the propagation constant is shifted into the gap of the photonic transmission spectrum. By employing a simple nonlinear discrete model~\cite{Kivshar:2003:OpticalSolitons}, we describe the crossover from discrete diffraction and surface repulsion in the linear regime, to the appearance of a purely nonlinear localized surface state at higher optical intensities. We discuss the physical mechanism of the nonlinearity-induced stabilization of the staggered surface modes.

\pict[0.99]{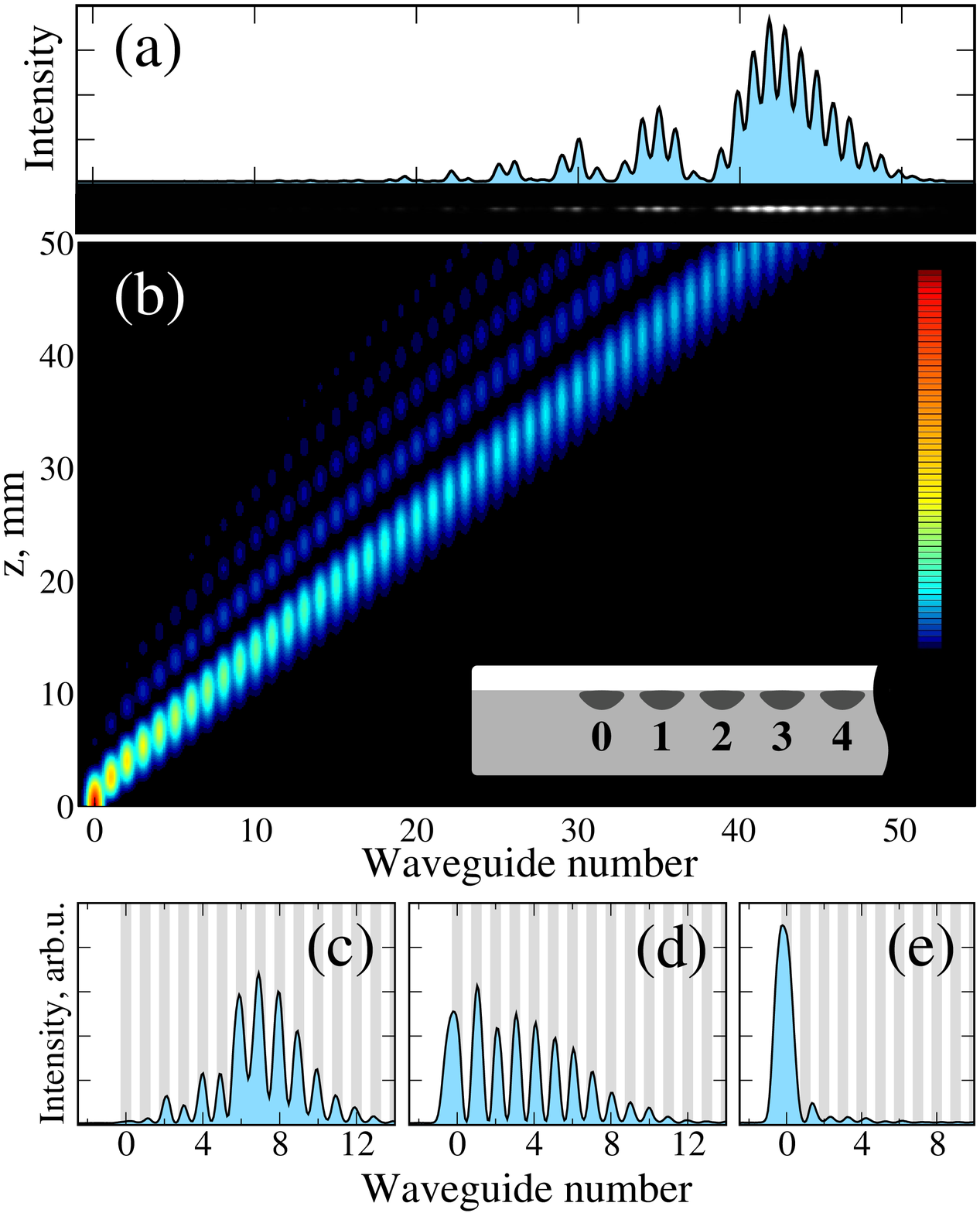}{lin-diffr}{(color online) Linear propagation of a narrow low-power beam when only the edge waveguide of the array is excited. (a)~Measured transverse output intensity profile ($P=0.1\,\mu$W) and (b)~corresponding theoretically calculated longitudinal propagation inside the sample. Inset in (b) shows the waveguide geometry. (c-e) Formation of the surface gap soliton at the array output 920, 1050, and $1550\,$s, respectively, after the input beam power is increased to $P=0.5\,$mW. Grey shading marks the waveguide positions.}

In our experiments, we study nonlinear surface localization in a semi-infinite array of single-mode optical waveguides fabricated by a Titanium in-diffusion process in a mono-crystal lithium niobate (LiNbO$_3$) wafer, similar to that recently used for the observation of discrete gap solitons~\cite{defoc_gap, Matuszewski:2006-254:OE}. The fabrication process, described in Ref.~\cite{Matuszewski:2006-254:OE},
results in a high-quality waveguide array with refractive index contrast $\Delta n=3\times10^{-4}$, waveguide spacing $d=9.0\,\mu$m, sample length $50\,$mm, and a total of 100 waveguides. Inset in Fig.~\rpict{lin-diffr}(b) shows schematically the geometry of the waveguide array. The LiNbO$_3$ sample exhibits a strong photovoltaic effect which leads to defocusing nonlinearity at visible wavelengths.

In the experimental setup an extraordinarily polarized probe beam from a cw Nd:YVO$_4$ laser ($\lambda=532\,$nm) is focused by a microscope objective ($\times20$) to a full width at half-maximum (FWHM) of $2.7\,\mu$m at the input face of the sample, and injected into the waveguide at the edge of the array. The propagated wavepacket at the output of the sample is imaged onto a CCD camera. The FWHM of the individual waveguide mode is $6\,\mu$m and $3\,\mu$m in horizontal and vertical directions, respectively, allowing for a single-waveguide input coupling.
The waveguide array is externally illuminated by a white-light source in order to control the nonlinear response time. As shown in Ref.~\cite{Matuszewski:2006-254:OE} single site excitation provides an efficient method for excitation of gap solitons in periodic defocusing nonlinear materials, provided the refractive index contrast exceeds a certain threshold. In this case the periodic structure appears equivalent to a discrete system~\cite{Matuszewski:2006-254:OE} and can be well described by a nonlinear discrete model.

At low laser power ($P=0.1\,\mu$W), we observe two major effects. First, due to coupling between neighboring waveguides the probe beam experiences discrete diffraction and spreads out in the horizontal plane upon propagation. Second, the beam shifts dramatically to the right indicating a strong repulsive effect of the surface. Figure~\rpict{lin-diffr}(a)  shows the experimental output image and the corresponding transverse intensity profile. After linear propagation through the array the beam profile acquires a complex form, spanning about 30 waveguides. The major lobe is centered approximately 42 lattice sites away from the input excitation point ($n=0$ at the edge of the array) due to the surface repulsion. Figure~\rpict{lin-diffr}(b) shows the corresponding optical intensity distribution inside the sample, calculated with the help of a simple analytical formula derived from a discrete model $a_n(z\kappa)=A_0i^n\,[J_n(2z\kappa)+J_{n+2}(2z\kappa)]$, where $a_n(z\kappa)$ is the discrete mode amplitude in the $n$-th waveguide, $A_0$ is the initial field amplitude in the input waveguide $n=0$, $z$ is the propagation distance, and $\kappa$ is the intersite coupling coefficient~\cite{Makris:2005-2466:OL}. In Fig.~\rpict{lin-diffr}(b) the discrete mode amplitudes have been multiplied by the continuous waveguide mode intensity profile, and the agreement with the experimental observation is found to be excellent. The coupling coefficient is estimated to be $\kappa=0.46\,$mm$^{-1}$, implying a total longitudinal propagation of 23 coupling lengths.

Increasing the laser power leads to spatial beam self-action through the defocusing photovoltaic nonlinearity. The slow response of the nonlinearity allows us to monitor directly the transient temporal dynamics of self-trapping and soliton formation, providing additional information about the localization process. Figures~\rpict{lin-diffr}(c-e) show the output beam intensity profile at times 920, 1050, and $1550\,$s, respectively, after the beam power is increased to $P=0.5$~mW. The wavepacket is seen first to contract and shift towards the edge of the array, indicating a nonlinearity-induced suppression of the surface repulsion [Fig.~\rpict{lin-diffr}(c)]. Then partial self-trapping at the surface occurs, with a tail of intensity lobes extending into the periodic structure [Fig.~\rpict{lin-diffr}(d)]. A series of zero intensity points between these lobes indicates the self-induced dynamic formation of {\em a staggered phase structure} which is clearly absent in Fig.~\rpict{lin-diffr}(c). Eventually, a strongly localized {\em surface gap soliton} is formed [Fig.~\rpict{lin-diffr}(e)]. The asymmetry of the photonic structure is reflected in the shape of the trapped beam which decays monotonically into the continuum while showing damped oscillations inside the array, resembling the structure of a {\em truncated Bloch mode}. The defocusing nonlinearity effectively decreases the contrast of the surface waveguide, causing the localized mode to broaden and penetrate substantially into the continuous medium.

\pict[0.9]{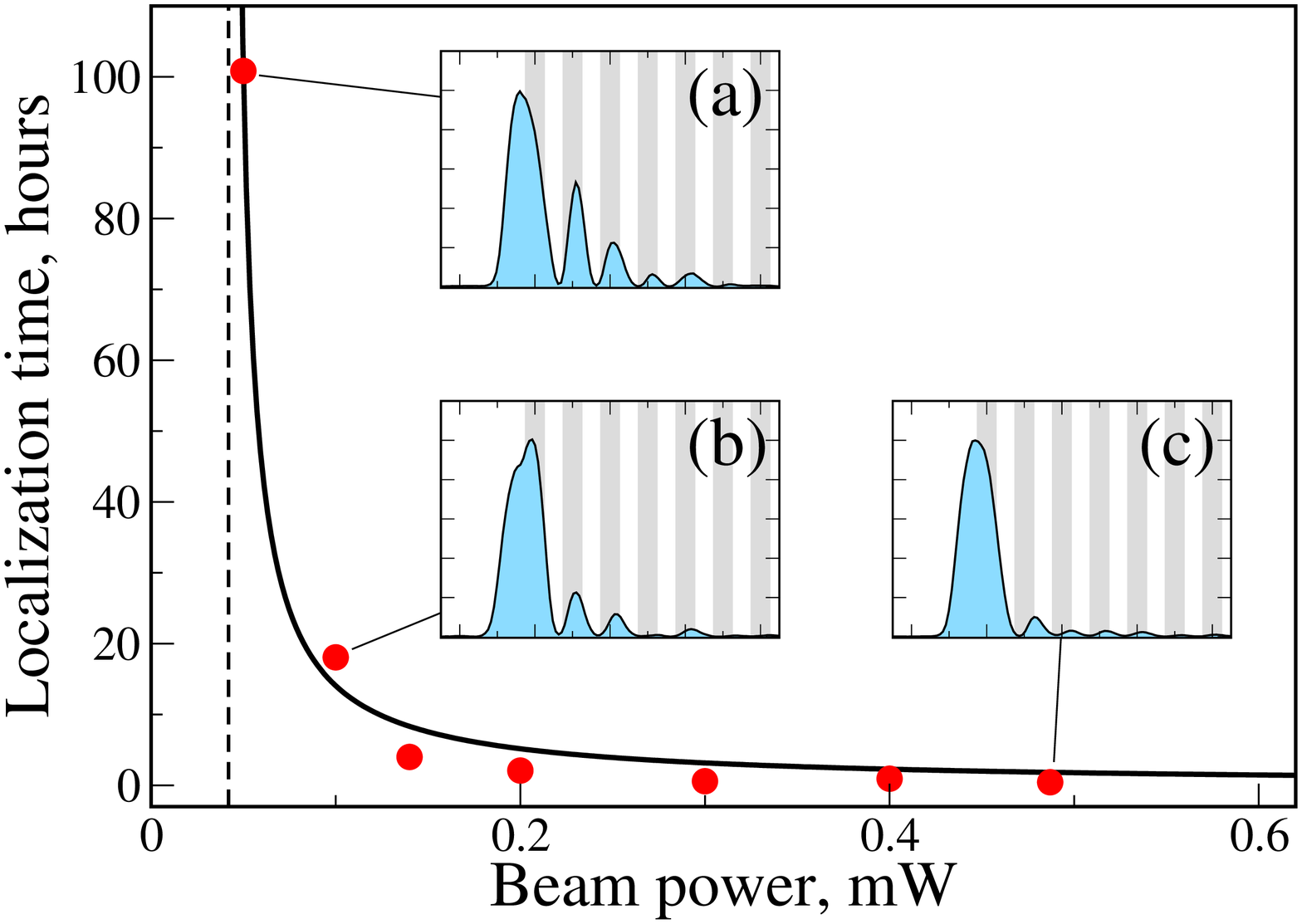}{threshold}{(color online)~Measured surface localization time vs. probe beam power. Solid curve: $A+B/(P-P_{\rm th})$ fit to experimental data (red dots). Vertical dashed line marks the threshold power ($P_{\rm th}=0.042\,$mW). (a-c)~Beam intensity profiles of decreasing width corresponding to the indicated points.}

In order to study in detail the crossover between linear diffraction and nonlinear self-localization, we measure the surface gap soliton formation time as a function of the probe beam power. The results are summarized in Fig.~\rpict{threshold}. The formation time increases dramatically for decreasing input power until, below a certain critical power, no localized surface mode is observed. The observed critical slowing down indicates the existence of a threshold power below which the nonlinear response is too weak to cause self-trapping. The value of the threshold power was estimated as $P_{\rm th}=0.042\,$mW by modelling the dynamics of the soliton formation time, fitting the function $A+B/(P-P_{\rm th})$ (Fig.~\rpict{threshold}, solid curve) to the experimental data (Fig.~\rpict{threshold}, red dots). Figures~\rpict{threshold}(a-c) show the beam intensity profiles corresponding to the indicated data points. The width of the localized mode decreases for increasing beam power, spanning about three lattice sites immediately above threshold [see Fig.~\rpict{threshold}(a)], and approximately a single lattice site at higher power, as indicated in Fig.~\rpict{threshold}(c). The decrease of the beam width is due to the fact that stronger beam self-action at higher power leads to a deeper surface defect, and hence more pronounced beam localization.

\pict[0.7]{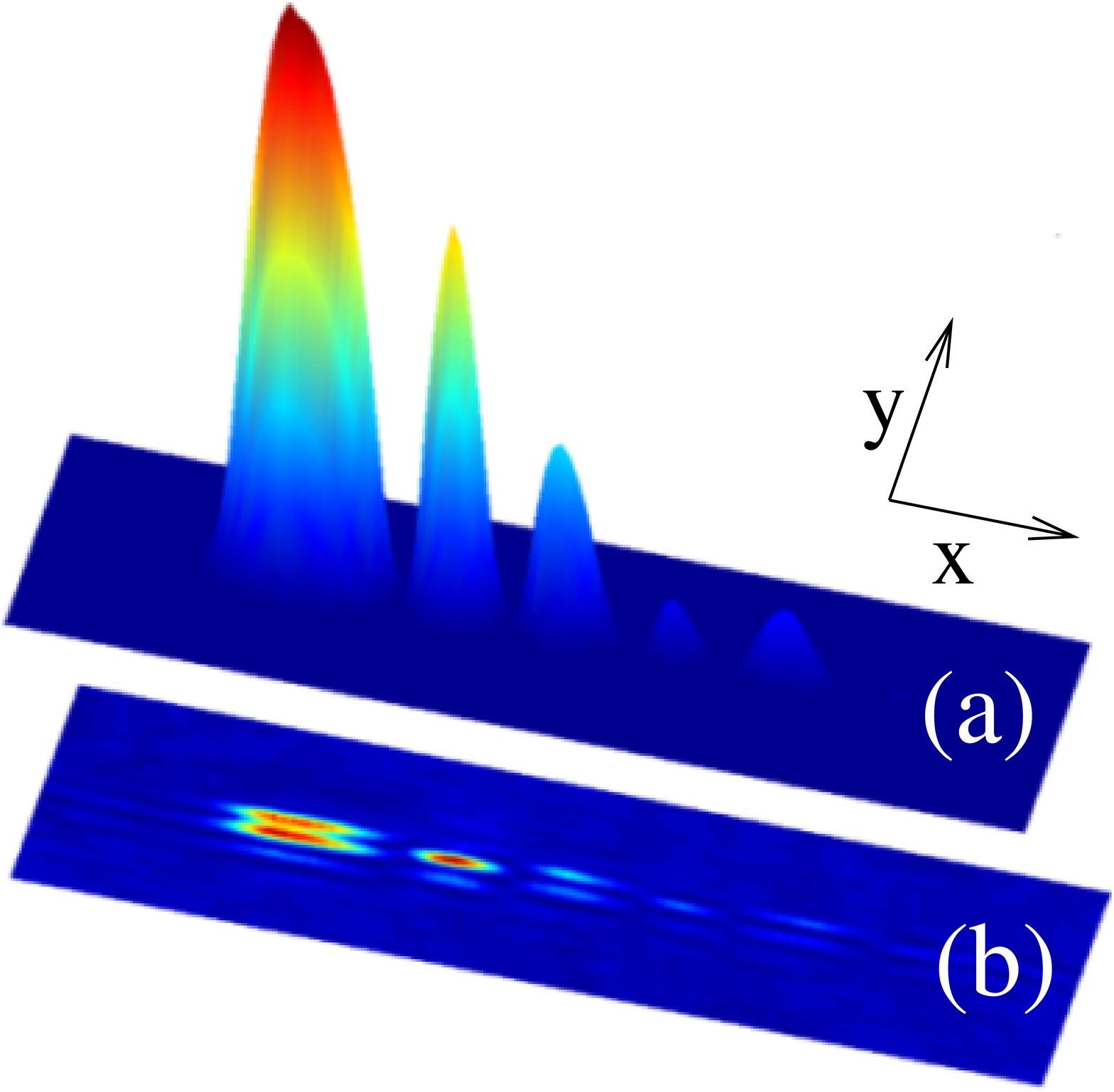}{peak-fringes}{(color online)~(a)~Three-dimensional representation of the surface gap soliton observed experimentally near the threshold [corresponding to Fig.~\rpict{threshold}(a)], (x,y) are the transverse coordinates. (b)~Plane-wave interferogram demonstrating the staggered phase structure of the surface gap soliton.}

An essential and unique feature of the observed surface gap solitons is the {\em staggered phase structure} of the beam tail inside the periodic medium. The alternating phase of the field lobes reflects the fact that the propagation constant of the self-localized mode lies within the photonic bandgap at the edge of the first Brillouin zone. To verify that this is indeed the case in the experiment, we interfere the output beam with a vertically inclined plane reference wave. Figure~\rpict{peak-fringes}(a) depicts a three-dimensional representation of the spatial beam intensity distribution of the broad surface gap soliton observed near the threshold [Fig.~\rpict{threshold}(a)]. Figure~\rpict{peak-fringes}(b) shows a two-dimensional intensity plot of the associated interference pattern.
A half-period vertical shift of the interference fringes, corresponding to an exact $\pi$ phase jump in the horizontal beam direction, is clearly observed between each pair of lobes in the structure [Fig.~\rpict{peak-fringes}(b)]. The phase is seen to be constant in the continuous region. The staggered phase structure inside the array and the plane phase in the continuum are signatures of the two different localization mechanisms in play~\cite{Kartashov:2006-073901:PRL}. The mode is confined from the continuum by total internal reflection, while Bragg reflection is responsible for localization inside the periodic structure.

In order to get a deeper insight into the physics of staggered surface soliton formation in semi-infinite lattices with defocusing nonlinearity, we consider the system of coupled-mode equations~\cite{discrete} for the normalized mode amplitudes $E_0$ and $E_n$ ($n = 1, 2, \ldots$), assuming weak coupling between the neighboring waveguides,
\begin{equation}
\label{eq:1}
   \begin{array}{l} {\displaystyle
i {d E_{0}\over{d z}} +  E_{1} + {\cal F}(E_{0}) E_{0}= 0,
   } \\*[9pt] {\displaystyle
i {d E_{n}\over{d z}} +  (E_{n+1} + E_{n-1}) + {\cal F}(E_{n}) E_{n}
=0,}
\end{array}
\end{equation}
where ${\cal F}(E) = \gamma /(1 +  |E|^2)$ accounts for the saturable character of the photovoltaic nonlinearity~\cite{model}. For a defocusing nonlinearity, $\gamma>0$.

Looking for stationary solutions in the form $E_{n}(z) = \exp(i\beta z) E_{n}$, we obtain the linear spectrum of extended modes, $\beta= 2 \cos \,k$, $(0\leq k \leq \pi)$. No localized surface mode exists in the linear regime, as this would require large index contrast between the waveguides and the continuum. However, the presence of defocusing nonlinearity in the model (\ref{eq:1}) can give rise to localized states. To find them we solve numerically the corresponding stationary equations by a multi-dimensional Newton-Raphson scheme. Since we are interested in surface localized modes, we look for states with maxima near the surface which decay quickly away from the edge of the array, similar to the earlier studied cases of Kerr nonlinearity for the discrete~\cite{Makris:2005-2466:OL} and continuous~\cite{Kartashov:2006-073901:PRL} models.

Figure~\rpict{theory-width}(a) shows the staggered surface mode calculated by use of the discrete nonlinear model (\ref{eq:1}) and multiplied by the waveguide mode-field profiles. Using the results solely based on the discrete model (\ref{eq:1}) provides a reasonable agreement with the experimental data. An example of such a comparison is shown in Fig.~\rpict{theory-width}(b) for the width of the localized surface state calculated numerically (solid curve) and measured experimentally (diamonds) as a function of the beam power.

\pict[1]{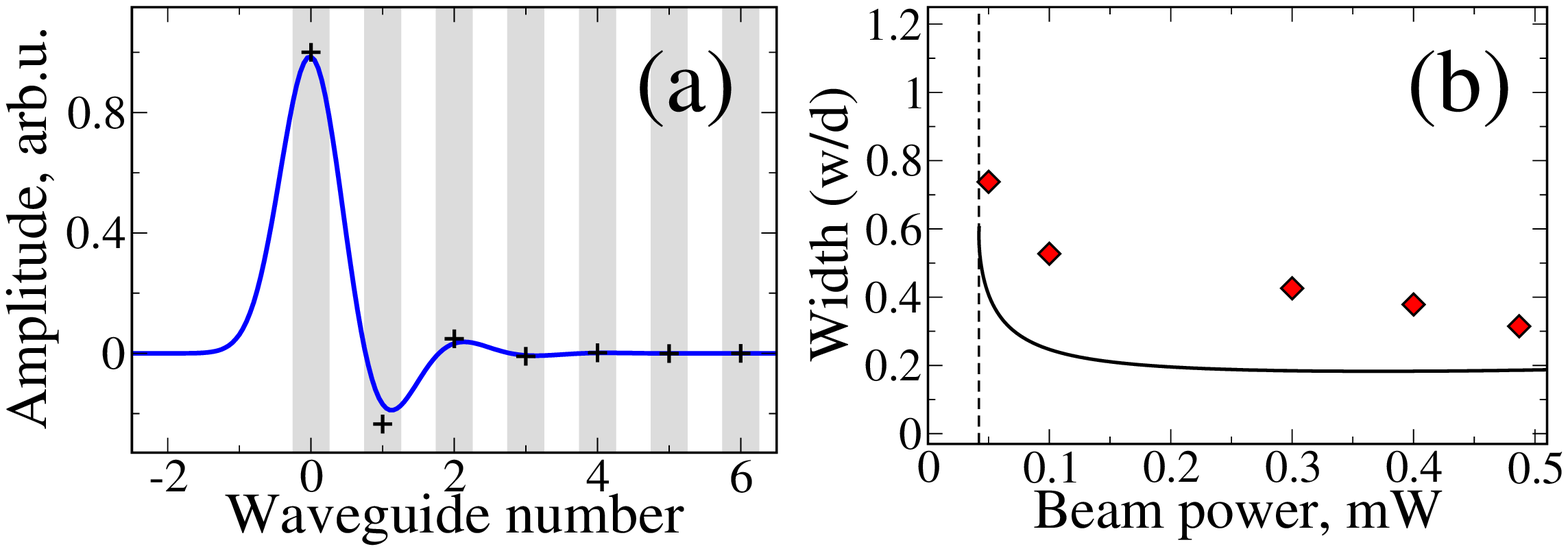}{theory-width}{(a) Example of the staggered surface mode calculated by use of the discrete nonlinear model (\ref{eq:1}) at $\gamma=8$. The discrete mode amplitudes are marked by +~signs. (b) Normalized width of the localized surface state calculated numerically (solid curve) and measured experimentally (diamonds) as a function of the beam power. Vertical dashed line marks the threshold power ($P_{\rm th}=0.042\,$mW).}

Despite being approximate, the discrete model can be employed to reveal an important physical mechanism of the nonlinearity-induced surface mode stabilization. To this end we follow earlier studies~\cite{kivshar,ol_mario} and calculate the effective energy of the mode, $H = -\sum_n \{(E_{n}E_{n+1}^{*} + E_{n}^{*}E_{n+1}) +\gamma \ln (1+ |E_{n}|^{2})\}$, as a function of its collective coordinate $X = P^{-1} \sum_n n|E_{n}|^2$, where $P = \sum_{n} |E_{n}|^2$ is the mode power. We apply a constraint method and start from the solution centered at the site $n=0$ for given values of $\beta$ and $P$. Our goal is to obtain all intermediate solutions between the neighboring stationary configurations for the same power. First, we calculate the stationary mode centered at $n=0$ and obtain all $\{E_{n}\}$ and the power $P$; then we fix the amplitude at the site $n=1$ to $E_{1} + \epsilon$, and solve the Newton-Raphson equations for all remaining $E_{m}$ ($m\neq 1$) with the constraint that the power be kept at $P$, arriving at an intermediate state centered between $n=0$ and $n=1$. Finally we vary $\epsilon$ and repeat the procedure until reaching the even configuration where $X=0.5$. The procedure is repeated for the solutions centered at $n=1,2,3$ which allows us to construct the effective potential plotted in Figs.~\rpict{theory-threshold}(a,b).

\pict[1]{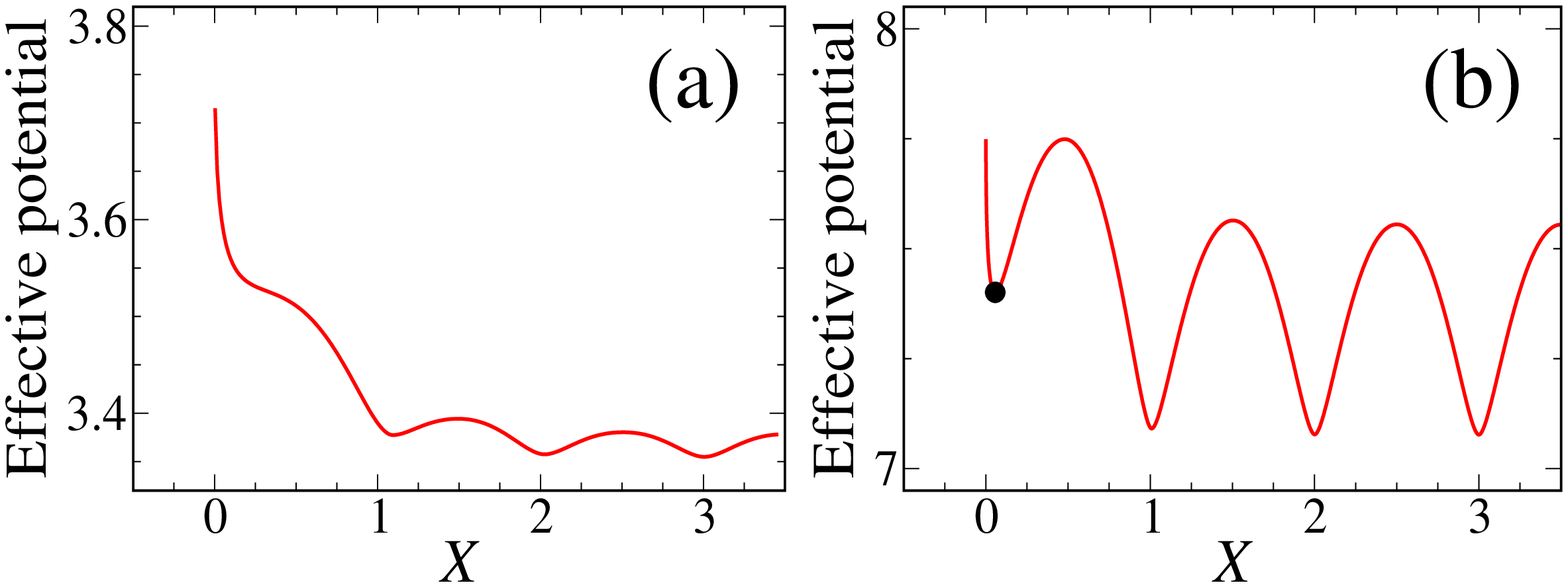}{theory-threshold}{(color online) (a,b) Effective potential of the localized gap modes vs. the collective coordinate $X$ below and above the threshold power, respectively. Integer values of $X$ correspond to the waveguide numbers. Black dot in (b) refers to the stationary solution shown in Fig.~4(a).}

Figures~\rpict{theory-threshold}(a,b) show the effective potential of the surface mode in a semi-infinite array, $U_{\rm eff}(X) \equiv - H(X)$, calculated for two different power values. The extremal points of this curve defined by the condition $dH/dX=0$ correspond to stationary localized solutions. In comparison with an infinite array, the truncation of the waveguide array introduces an effective {\em repulsive surface potential}, which is combined with the periodic potential of the array. As a result, discrete surface modes are possible neither in the linear regime nor in the continuous limit. As we see from Fig.~\rpict{theory-threshold}(a), for low powers there exists no solution of the equation $dH/dX=0$ at the surface site $n=0$, and the surface repels the input beam as clearly observed in experiment [see Fig.~\rpict{lin-diffr}(a,b)]. However, when the input power exceeds the threshold value, discreteness overcomes the surface repulsive force and the localized state at $n=0$ in the form of the surface gap soliton becomes possible [Fig.~\rpict{theory-threshold}(b)].

In conclusion, we have predicted theoretically and demonstrated experimentally that gap solitons can be stabilized near the surface of a periodic medium with self-defocusing nonlinearity in the form of staggered surface modes, providing the first experimental evidence of a nonlinear analog of surface Tamm states in optics.

The authors thank A.A. Sukhorukov for discussions and S. Flach for a help with the constraint method, and acknowledge a support from Fondecyt (grants 1050193 and 7050173) and the Australian Research Council.


\end{sloppy}

\begin{thebibliography}{10}

\bibitem{Tamm:1932-849:ZPhys} I.~E. Tamm, Z. Phys. {\bf 76}, 849 (1932).

\bibitem{Ohno:1990:PRL} H. Ohno {\em et al.}, Phys. Rev. Lett. {\bf 64}, 2555 (1990).

\bibitem{yariv}  P. Yeh, A. Yariv, and C.-S. Hong, J. Opt. Soc. Am. {\bf 67}, 423 (1977).

\bibitem{Yeh_APL_78} P. Yeh, A. Yariv, and A.~Y. Cho, Appl. Phys. Lett. {\bf 32}, 102 (1978).

\bibitem{tomlinson} W.~J. Tomlinson, Opt. Lett. {\bf 5}, 323 (1980).

\bibitem{book} A.~D. Boardman, {\em et al.}
in: {\em Nonlinear Surface Electromagnetic Phenomena}, Eds. H.~E. Ponath and G.~I. Stegeman (North-Holland, Amsterdam, 1991), Vol. 29, p. 73.

\bibitem{mihalache} D. Mihalache, M. Bertolotti, and C. Sibilia, Prog. Opt. {\bf 27}, 229 (1989).

\bibitem{physica_d} Yu.~S. Kivshar, F. Zhang, and S. Takeno, Physica D {\bf 113}, 248 (1998).

\bibitem{Makris:2005-2466:OL} K.~G. Makris {\em et al.} Opt. Lett. {\bf 30}, 2466 (2005).

\bibitem{Suntsov:2006-063901:PRL} S. Suntsov {\em et al.}, Phys. Rev. Lett. {\bf 96}, 063901 (2006).

\bibitem{Kartashov:2006-073901:PRL} Ya.~V. Kartashov, V.~V. Vysloukh, and L. Torner,
Phys. Rev. Lett. {\bf 96}, 073901 (2006).

\bibitem{Kivshar:2003:OpticalSolitons}
Yu.~S. Kivshar and G.~P. Agrawal, {\em {Optical Solitons: From Fibers to Photonic Crystals}} (Academic Press, 2003).

\bibitem{defoc_gap} F. Chen {\em et al.}, Opt. Expr. {\bf 13}, 4314 (2005).

\bibitem{Matuszewski:2006-254:OE}
M. Matuszewski {\em et al.}, Opt. Expr. {\bf 14}, 254 (2006).

\bibitem{discrete} D.~N. Christodoulides and R.~I. Joseph, Opt. Lett. {\bf 13}, 794 (1988);
Yu.~S. Kivshar, Opt. Lett. {\bf 18}, 1147 (1993).

\bibitem{model} Lj. Had{\u z}ievski {\em et al.}, Phys. Rev. Lett. {\bf 93}, 033901 (2004).

\bibitem{kivshar} Yu.~S. Kivshar, F. Zhang, and A.~S. Kovalev, Phys. Rev. B {\bf  55}, 14265 (1997).

\bibitem{ol_mario} M. Molina, R. Vicencio, and Yu.~S. Kivshar, Opt. Lett. {\bf 31} (2006) in press.

\end{thebibliography}
\end{document}